\documentclass[preprint,12pt]{aastex}           
\usepackage{natbib,graphics}

\newcommand{\al}{\alpha}
\newcommand{\be}{\begin{equation}}
\newcommand{\ee}{\end{equation}}
\newcommand{\bdm}{\begin{displaymath}}
\newcommand{\edm}{\end{displaymath}}
\newcommand{\bea}{\begin{eqnarray}}
\newcommand{\eea}{\end{eqnarray}}

\newcommand{\Sc}{{\cal S}}
\newcommand{\Om}{\Omega}
\newcommand{\Omt}{\tilde{\Omega}}
\newcommand{\Jd}{\dot{J}_a}
\newcommand{\Md}{\dot{M}}

\newcommand{\Te}{T_8}

\begin{document}

\title{Conditions for Steady Gravitational Radiation from Accreting Neutron Stars} 

\author{Robert V. Wagoner\altaffilmark{1}} 
\affil{Dept. of Physics and Center for Space Science and Astrophysics \\ 
Stanford University, Stanford, CA 94305--4060}
\altaffiltext{1}{wagoner@stanford.edu} 

\begin{abstract}
The gravitational-wave and accretion driven evolution of the angular velocity, core temperature, and (small) amplitude of an r-mode of neutron stars in low mass X-ray binaries and similar systems is investigated. The conditions required for evolution to a stable equilibrium state (with gravitational wave flux proportional to average X-ray flux) are determined. In keeping with conclusions derived from observations of neutron star cooling, the core neutrons are taken to be normal while the core protons and hyperons and the crust neutrons are taken to be singlet superfluids. The dominant sources of damping are then hyperon bulk viscosity (if much of the core is at least 2--3 times nuclear density) and ($e$-$e$ and $n$-$n$) shear (and possibly magnetic) viscosity within the core--crust boundary layer. It is found that a stable equilibrium state can be reached if the superfluid transition temperature of the hyperons is sufficiently small ($\lesssim 2\times 10^9$ K), allowing the gravitational radiation from Sco X-1 and several other neutron stars in low-mass X-ray binaries to be potentially detectable by the second generation LIGO (and VIRGO) arrays.  
\end{abstract}

\keywords{accretion, accretion disks --- dense matter --- gravitational waves --- stars: neutron --- X-rays: binaries}

\section{Introduction}

In this Letter, we present the results of an investigation of the evolution of rapidly rotating accreting neutron stars under the influence of their emission of gravitational radiation. We employ a modification \citep{whl} and extension of the two-component (equilibrium plus perturbation) model of the star introduced by \citet{owen98} and also employed by \citet{lev99}, but restrict the analysis to small perturbations. These are assumed to be in the form of r-modes \citep{and98,fm98,lom98,aks99}, which radiate mainly via Coriolis-driven velocity perturbations rather than the density perturbations of the less powerful f-modes. We must allow for large uncertainties in many of the relevant properties of neutron stars, such as the superfluid transition temperatures and the properties of the core-crust boundary layer. Details will be presented in a subsequent paper. 

After developing a general formalism, we shall focus on conditions in which the neutron star angular velocity (and thus gravitational wave frequency) evolves slowly toward an equilibrium state, in which the rate of accretion of angular momentum from the surrounding disk is balanced by its rate of loss via gravitational radiation. If this equilibrium is achieved, the observed flux of gravitational radiation can be shown to be proportional to the observed flux of X-rays from the accretion \citep{wag84,bil98}. 

One of our longer term goals is the development of parameterized expressions describing possible time evolutions of the gravitational-wave frequency and amplitude, to facilitate detection by LIGO, VIRGO, and similar laser interferometer detectors. The brightest low mass X-ray binaries (LMXBs) thought to contain a neutron star are the prime targets.

\section{Dynamical and Thermal Evolution}

In this exploratory investigation, it will be sufficient to consider a Newtonian neutron star in equilibrium (with equatorial radius $R$) which is perturbed by a nonaxisymmetric infinitesimal fluid displacement $\vec{\xi}=\vec{f}(r,\theta)e^{i(m\phi +\sigma t)}\sim \alpha R$, with $\al\ll 1$.
Based on the work of \citet{fs78a} and \citet{lu00}, the total angular momentum $J$ of the star can be decomposed into its equilibrium angular momentum $J_*$ and a perturbation proportional to the canonical angular momentum $J_c$. That is,
\be
J = J_*(M,\Om) + (1-K_j)J_c \; , \qquad J_c = -K_c\al^2 J_* \; , \label{decomp}
\ee
where $M$ is the mass and $\Om$ is the (uniform) angular velocity of the equilibrium star. All constants $K_{(\;)}$ will be dimensionless, with $K_j\sim K_c\sim 1$.

In classical mechanics, the action $I=E/\omega$ of any normal mode of a set of oscillators (with frequency $\omega$) is an adiabatic invariant. For a fluid, the analogous quantity should be $\tilde{E}_c/\omega$, where $\tilde{E}_c$ is the canonical energy of the perturbation in the corotating frame and $\omega=\sigma+m\Omega$ is its frequency in that frame. However, we also have the general relation $\tilde{E}_c = -(\omega/m)J_c$ \citep{fs78a}. Therefore, following \citet{hl00}, we assume that the canonical angular momentum is also an adiabatic invariant, and should therefore be unaffected by the slow rate of mass accretion. Thus it obeys the usual relation \citep{fs78b}
\be 
dJ_c/dt = 2J_c[(F_g(M,\Om)-F_v(M,\Om,T_v)] \; ,                  \label{canonical}
\ee
where $F_g$ is the gravitational radiation growth rate and $F_v$ is the viscous damping rate. The latter also depends upon a spatially averaged temperature $T_v(t)$. 

On the other hand, conservation of total angular momentum requires that
\be
 dJ/dt = 2J_c F_g + \Jd(t) \; ,                               \label{totangmom}
\ee
where $\Jd=j_a\Md$ is the rate of accretion of angular momentum. The mass is accreted with specific angular momentum $j_a$ at a rate $\Md(t)$. 

Combining these equations then gives the dynamical evolution relations
\bea
{1\over\al}{d\al\over dt} & = & F_g-F_v + [K_jF_g+(1-K_j)F_v]K_c\al^2 - \left({j_a\over 2J_*}\right)\Md(t) \; , \label{dadt} \\
\left({I_*\over J_*}\right){d\Om\over dt} & = & -2[K_jF_g+(1-K_j)F_v]K_c\al^2 + \left[{(j_a-j_*)\over J_*}\right]\Md(t) \; ; \label{dodt}
\eea
where $I_*(M,\Om)\equiv\partial J_*/\partial\Om$ and $j_*(M,\Om)\equiv\partial J_*/\partial M$.
In keeping with the fact that it is sufficient to also work to lowest order in $\Om/\Om(\mbox{max})$ (as well as the relativity parameter $GM/Rc^2$) in this exploratory investigation, we take $j_a-j_*\cong j_a$ and $J_*\cong I_*\Om$. We also note that $K_c \cong 0.094$ \citep{owen98}. (The value of $K_j$ is unimportant, since we will see that $F_v$ remains very close to $F_g$.) 

Finally, thermal energy conservation for the entire star gives our third evolution equation
\be
\int{\partial T\over\partial t}c_v dV \equiv C(T){dT\over dt} \cong 2\tilde{E}_c F_v(T_v) + K_n\langle\Md\rangle c^2 - L_\nu(T_\nu) \; , \label{entire}
\ee
where the corotating frame canonical energy ${\tilde E}_c = K_e\Om J_*\al^2$, with $K_e=K_c/3$. 
Since all constituents (electrons, nucleons,$\ldots$) are degenerate, the specific heat at constant volume ($c_v$) is essentially the same as that at constant pressure. The terms on the right hand side of this equation represent viscous heating, pycnonuclear reactions and neutron emissions in the inner crust (proportional to a time-averaged mass accretion rate), and neutrino luminosity. The hydrogen/helium burning rate is assumed to be balanced by the surface emission of photons \citep{sch99}, especially at the large accretion rate $\Md = 10^{-8}M_\odot\mbox{ yr}^{-1}$ (roughly 1/3 the Eddington rate, appropriate to our primary targets) that we shall adopt. 
The mass accretion rate can be estimated from accretion energy conservation. The photon luminosity arising directly from the accretion is $L_{acc} \approx (GM/R)\Md(t)$, for a slowly rotating neutron star with a negligible magnetosphere.

From now on we shall take the perturbation to be due to the dominant $l=m=2$ r-mode, in which case the gravitational wave frequency $f_{gw} = (4/3)f_{ns} = 2\Om/3\pi$. In order to facilitate comparison with previous results, we shall adopt the neutron star model of \citet{owen98} ($p\propto\rho^2$, $M=1.4M_\sun$, and $\Om_c\equiv (\pi G\langle\rho\rangle)^{1/2} = 8.1\times 10^3$ rad/s) for numerical work. Then the gravitational radiation growth rate of this mode is 
\be 
F_g = \Omt^6/\tau_{gr} \; , \quad \tau_{gr} = 3.26 \mbox{ s} \; , 
\qquad \Omt\equiv\Om/\Om_c \; , \label{Fg}
\ee
which should also dominate that due to the f-modes. 

Comparison of observations of thermal emission from isolated neutron stars with computed cooling histories \citep{tsu} has led \citet{kyg} to conclusions regarding the most likely state of the  constituents, which we tentatively adopt. Specifically, the maximum values of the (density dependent) superfluid transition temperatures are taken to be (a) $T_n\la 10^8$ K for the (triplet) core neutrons (so they will remain normal), (b) $T_p\ga 5\times 10^9$ K for the (singlet) core protons, and (c) $T_n\ga 5\times 10^9$ K for the (singlet) inner crust neutrons. In what follows we shall also assume that the thermal conductivity timescales $\tau_{th}$ are short enough to give relations $T_v(T)$ and $T_\nu(T)$ between these three spatially averaged temperatures that appear in equation (\ref{entire}). Typically, $\tau_{th}\sim 0.1-1$ year in the core and $\tau_{th}\sim 10-10^2$ years in the inner crust \citep{br00,bhy}.  
   
Then in the temperature range of interest ($10^8\la T \la 10^9$ K), the viscous damping rate of this mode is
\be
F_v \cong F_{sh}(T) + F_{bl}(\Om,T) + F_{hb}(\Om,T)  \; . \label{Fv}
\ee
Even if the neutrons were not normal, the contribution to this damping rate from the mutual friction between a neutron superfluid and the superconducting proton---relativistic electron fluid \citep{lm00} would be negligible.

The first term is produced by the ordinary shear viscosity throughout the core. Adding the contribution of the $n$-$n$ scattering \citep{owen98} to that of the $e$-$e$ scattering \citep{lm00} gives
\be
F_{sh} = 1/(\tau_{sh}\Te^2) \; , \quad \tau_{sh}\cong 0.69\times 10^6\mbox{ s} \; , 
\quad \Te\equiv T/10^8\mbox{ K} \; . \label{sh}
\ee

The second term is produced by the ordinary shear and magnetic viscosity in the crust--core boundary layer. For a normal fluid, this damping rate has been calculated by \citet{ajks}, \citet{bu00}, \citet{lou00}, \citet{rie}, and \citet{wu} neglecting magnetic effects, while \citet{bu00}, \citet{m01}, and \citet{km} included them. 
For superfluid neutrons and protons, this rate was calculated by \citet{km}. We shall approximate the damping rate for our case (normal neutrons and superfluid protons in the boundary layer) by adding the rate of \citet{km} for a normal uncharged fluid to their superfluid rate when only the proton vortices (magnetic flux tubes) are pinned to the crust. The magnetic damping is due to the interaction of the field with the effective current density of the proton vortices, producing cyclotron-vortex waves. Although there is no mutual friction because no neutron vortices are present, we also neglect the small coupling between these fluids produced by $n$--$e$ scattering. We then obtain (after an approximate algebraic simplification)   
\be
F_{bl} = 3.15\times 10^{-4}\Sc_{ns}^2\sqrt{\Om}/\Te 
       + 3.71\times 10^{-8}\Sc_{s}^2 \sqrt{B} \; \mbox{ s}^{-1}            \; , \label{bl}
\ee
where $B$ (Gauss) is the radial magnetic field. The slippage factor $\Sc_s$ is the fractional degree of pinning of the vortices in the crust \citep{km}; and $\Sc_{ns}^2\equiv (2\Sc_n^2+\Sc_s^2)/3$, with the slippage factor $\Sc_n$ the fractional difference in velocity of the normal fluid between the crust and the core \citep{lu01}. Although both slippage factors were defined to be at most unity, we also let them contain the uncertainties in our model. [If the pinning were dominated by neutron vortices, the coefficients in equation (\ref{bl}) would be $(2-6)\times 10^3$ times larger, resulting in no growth of the mode unless $\Sc_s\ll 1$.] In obtaining most of the numerical results below, we take the magnetic contribution in equation (\ref{bl}) to be negligible, which requires that $B\la 6\times 10^{10}[1+2(\Sc_n/\Sc_s)^2)]^2\Omt\Te^{-2}$ G. 

The third term arises from the bulk viscosity produced by out--of--equilibrium hyperon reactions (which dominate that produced by direct and modified Urca reactions). This has been studied by \citet{j01a,j01b} and \citet{lo02} for normal nuclear matter and by \citet{hly} for superfluids. We employ the results of \citet{lo02} for $n+n\rightleftarrows p+\Sigma^-$. However, with our assumptions about superfluidity, the reaction whose rate is least reduced by superfluid phase space blocking (but is more difficult to calculate) should be $n+n\rightleftarrows n+\Lambda$. These hyperons should be present at densities $\rho\ga (5-8)\times 10^{14}\mbox{ g cm}^{-3}$, which is achieved over a large fraction of the core of relevant mass [$M\cong (1.3-1.6)M_\sun$] neutron star models for many nuclear equations of state (which are softened by their presence) \citep{blc}. Employing the superfluid reduction factor $R_{hb}(T/T_h)$ of \citet{hly} for normal nucleons and a hyperon with singlet superfluid transition temperature $T_h$, we obtain
\be
F_{hb}=f_{hb}t_0^{-2}\tau(T)/[1+(2\Om\tau(T)/3)^2]\; ,\quad 
\tau(T)= t_1 \Te^{-2}/R_{hb}(T/T_h) \; , \label{hb}
\ee
where $1/\tau$ is the reaction relaxation rate. The constants $t_0\approx 0.15$ s and $t_1\approx 0.01$ s are obtained by fitting the results of \citet{lo02}. The factor $f_{hb}$ allows for the difference in the rate of the two reactions and the fraction of the neutron star that has a density high enough to allow the reactions. Since the dominant uncertainty in equation(\ref{hb}) is due to that of $T_h$, we shall set $f_{hb}=1$, implying that the neutron star is sufficiently massive to have a central density well above the indicated hyperon threshold. 

The temperature dependences of these contributions to the viscous damping rate are shown in Figure \ref{Fv}.
Now that we have specified all properties in the equations (\ref{dadt}) and (\ref{dodt}) of evolution of $\al(t)$ and $\Om(t)$, we can consider the thermal evolution [equation(\ref{entire})]. With the neutrons normal, they dominate the specific heat, giving $C(T)\cong 1.5\times 10^{38}\Te$ erg/K. (The electron contribution is about 15 times less.) We also take the nuclear heating constant $K_n = 1\times 10^{-3}$ \citep{br00}.

The neutrino luminosity is taken to be
\be
L_\nu = L_{du}\Te^{\,6} R_{du}(T/T_p) + L_{mu}\Te^{\,8} R_{mu}(T/T_p) + L_{ei}\Te^{\,6} + L_{nn}\Te^{\,8} + L_{cp}\Te^{\,7} \; . \label{nu}
\ee
The proton superfluid reduction factors for the direct Urca reactions ($R_{du}$) and the modified Urca reactions ($R_{mu}$), as well as most of the following constants, are obtained from the review of \citet{yls}. The other terms represent (inner crust) electron--ion and (core) neutron--neutron neutrino bremsstrahlung, and Cooper pairing of (inner crust) neutrons. We take $L_{du}=f_{du}\times 10^8 L_{mu}$, where $f_{du}$ includes the fraction of the neutron star that is above the density threshold for the direct Urca reactions (comparable to the above threshold for the hyperon reactions). The constants $L_{mu}\approx 1.0\times 10^{32}$ erg/s and $L_{ei}\approx 9.1\times 10^{29}$ erg/s are obtained by fitting the results of \citet{br00} (who did not consider the other processes) for normal and superfluid neutron stars. Finally, $L_{nn}\approx 0.01 L_{mu}$, while $L_{cp}\approx 9\times 10^{31}$ erg/s is proportional to the length scale of variation of the neutron superfluid transition temperature within the inner crust  (taken to be 100 m) \citep{ykg}.

We are interested in the evolution of neutron stars after they have been spun up to the point where the gravitational radiation growth rate has become equal to the viscous damping rate: 
\be
F_g(\Om_0,M_0)=F_v(\Om_0,M_0,T_0)\equiv F_0 \; .        \label{initial}
\ee
This equality defines our initial state. Before that time, we see from equation(\ref{canonical}) that any intrinsic perturbation could not grow from its (infinitesimal) value $\al_{min}$.
The initial temperature $T_0$ is then determined by the vanishing of equation (\ref{entire}), with the nuclear heating in the inner crust balanced by the neutrino emission \citep{br00}. This  temperature is most sensitive to the direct Urca factor $f_{du}$, but only varies by $14\%$ over the range $0\leq f_{du}\leq 1$. Consistent with our assumption that the hyperon reactions are operative, we shall adopt $f_{du}=0.1$. 

For a typical value $\Omt_0\sim 0.3$, the time scale $\tau_0\equiv 1/F_0\sim 10^4$ sec. Two other key time scales are that due to cooling and accretion, 
\be
\frac{1}{\tau_c}\equiv F_c\equiv\frac{L_\nu(T_0)}{C(T_0)T_0}\sim \frac{1}{10^3\mbox{ yr}}\; , \qquad \frac{1}{\tau_a}\equiv F_a\equiv \left(\frac{j_a}{J_0}\right)\langle\Md\rangle\sim\frac{1}{10^7\mbox{ yr}} \; .
\ee
Since we will be concerned with time scales $\Delta t\ll M_0/\langle\Md\rangle\ga 10^8$ yr, we can take $M(t)\cong M_0$.   

In contrast to the initial state, the equilibrium state $X^i_e$ of our dynamical variables 
$X^i(t)=\{\al,\Om,T\}$ is defined by the vanishing of the evolution equation (\ref{totangmom}), in addition to equations (\ref{canonical}) and (\ref{entire}). From equation (\ref{dadt}) or (\ref{dodt}), one can see that the equilibrium amplitude is given by
\be
\al_e=\left[\frac{F_a}{2K_cF_g(\Om_e)}\right]^{1/2} \sim (10^{-6}-10^{-5}) \; . \label{aleq}
\ee
These values of $\al_e$ are much less than the saturation amplitude of the r-mode instability \citep{afm}. 

The linearized analysis of \citet{whl} shows that stability of the equilibrium requires that
\be
\left(\frac{1}{L_\nu}\frac{\partial L_\nu}{\partial T}\right)_e >
\left(\frac{1}{F_v}\frac{\partial F_v}{\partial T}\right)_e > 0 \; , \label{st}
\ee
assuming that $|\partial/\partial T|\sim 1/T$. If so, oscillations of frequency $f_o\sim K_r^{1/2}\al_eF_e$ are damped at a rate $f_d\sim K_r\al_e^2F_e$, where $F_e$ is the equilibrium value of $F_v=F_g$ and $K_r \equiv 2K_e\Om_eJ_0/C_eT_e\sim 10^5$ is the ratio of rotational to thermal energy. We have also used the fact that the viscous heating term in equation (\ref{entire}) is typically at least ten times greater than the nuclear heating term. In practice, the first inequality in equation (\ref{st}) is always satisfied, if the key requirement $\partial F_v/\partial T > 0$ holds (which is the case if $F_v$ is dominated by the hyperon bulk viscosity, as shown in Figure \ref{Fv}).
This requirement applied to the initial state also guarantees slow evolution, as we shall see below.

\section{Conclusions}

In Figure \ref{Omt} we show the critical curve $\Omt(\Te)$ given by $F_g=F_v$, for three choices of the key parameters $T_h,\Sc_{ns}$. Also shown are the initial state $\Omt_0, T_0/10^8$K and the equilibrium state $\Omt_e, T_e/10^8$K, with $T_e > T_0=3.32\times10^8$ K. For $T_h\ga 3\times 10^9$ K, these results are independent of $T_h$. For $T_h\la 1\times 10^9$ K, these results are independent of the slippage factors $\Sc_n$ and $\Sc_s$ (and $\Omt_e$ does not increase greatly as $T_h$ decreases in this range).

\begin{figure}
\plotone{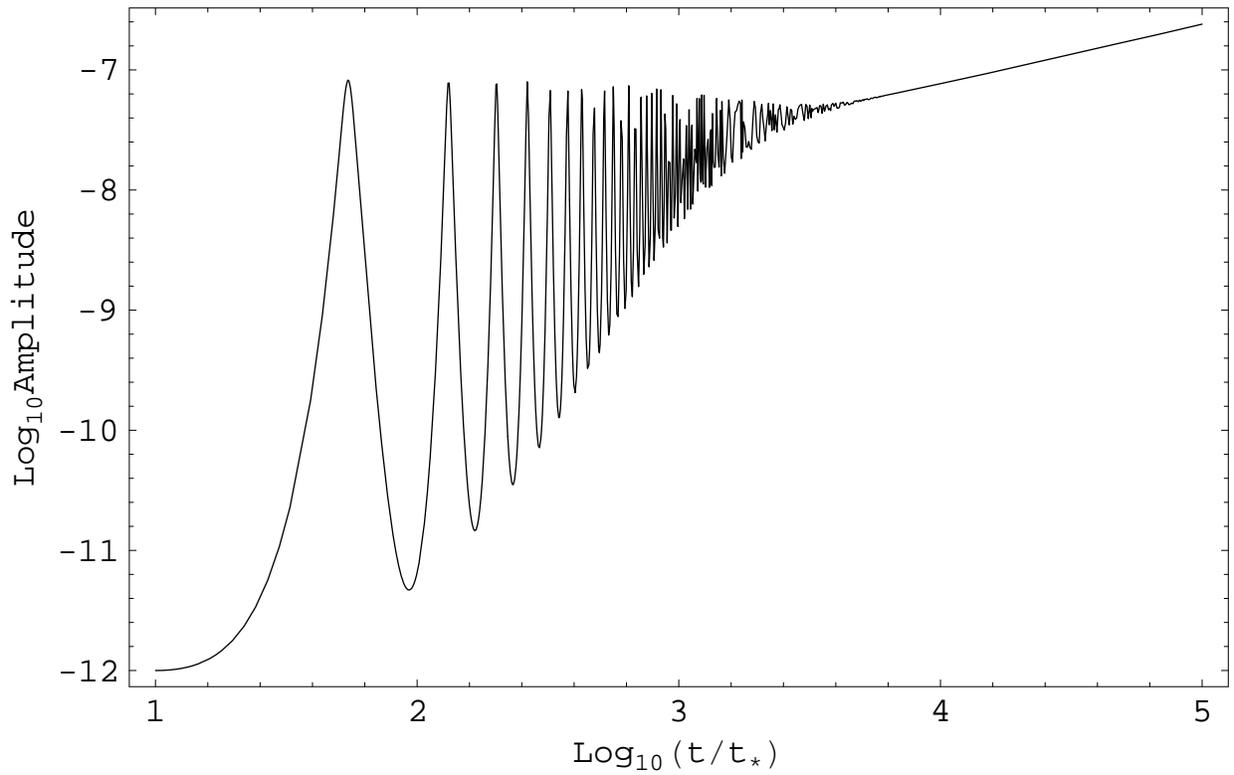}
\figcaption[Amp-l.eps]{The early evolution of the r-mode amplitude $\al$, for case (c) in Figure 2. The initial amplitude was chosen to be $\al_0=10^{-12}$, and $t_*=1$ year.\label{Amp}}
\end{figure}

It can be shown that the evolution from the initial state is also controlled by the sign of $(\partial F_v/\partial T)_0$, which is equal to the sign of the slope of the critical curve. If the slope is negative [case (a)], there will be a thermal runaway with a growth rate $f_o$ (given above) that is of the same magnitude as found by \citet{lev99}. If the slope is close to zero [case (b)], there will initially be overstable oscillations of the type found by \citet{whl}. If the slope is positive [case (c)], the oscillations of the growing amplitude are damped out on a timescale $\tau_c$ (as shown in Figure \ref{Amp}), after which it slowly increases to its equilibrium value [$\al_e=1.9\times 10^{-6}$ for the parameters of case (c)], along with $\Om$ and $T$. The time required to reach equilibrium is $\Delta t\approx (\Delta\Om/\Om)\tau_a$. Throughout, $F_g$ remains very close to $F_v$, so the evolution is along the critical curve. 

The maximum rotation rate (due to shedding) of our chosen neutron star model is $f_{max}\cong (2/3)(\Om_c/2\pi)=856$ Hz. Coherent oscillations in Type 1 X-ray bursts have been observed at frequencies $F\leq  590$ Hz \citep{vdk}, which would then correspond to $\Omt\lesssim 0.46$ if they represented the rotation rate. There is evidence that some of these are the first harmonic, in which case the highest spin frequency is 350 Hz \citep{vdk}. (For isolated neutron stars, $f\leq 642$ Hz.)

Observations of the luminosity of LMXB's in their quiescent (low accretion rate) phase constrain the r-mode amplitude if they are in the equilibrium state we have considered \citep{bru00}, where the r-mode viscous heating balances the neutrino energy loss. This will also be addressed in a subsequent investigation.

Our main conclusion is that evolution to a stable equilibrium state can occur if (a) a significant fraction of the core of the neutron star is above the threshold for hyperons, (b) their superfluid transition temperature $T_h\la 2\times 10^9$ K, (c) the core neutrons near the crust are not a superfluid whose vortices are strongly pinned to the crust, and (d) the magnetic field is not too strong in that core--crust boundary layer. If Sco X-1 has been spun up by accretion to such a stable equilibrium state, it should be detectable by the second generation LIGO detectors. (However, its spin period remains unknown.) When signal recycling (`narrow-banding') is employed, a few additional LMXB's may also be detectable \citep{ct}. 

\acknowledgments

   This work was supported in part by NSF grant PHY--0070935. We benefitted from many discussions during the 2000 program on Spin and Magnetism in Young Neutron Stars at the Institute for Theoretical Physics, U.C. Santa Barbara. Joseph Hennawi and Jingsong Liu provided important input to an early stage of this investigation. Thanks also goes to Lee Lindblom and Greg Mendell for very helpful discussions.

\newpage

\begin{figure}
\plotone{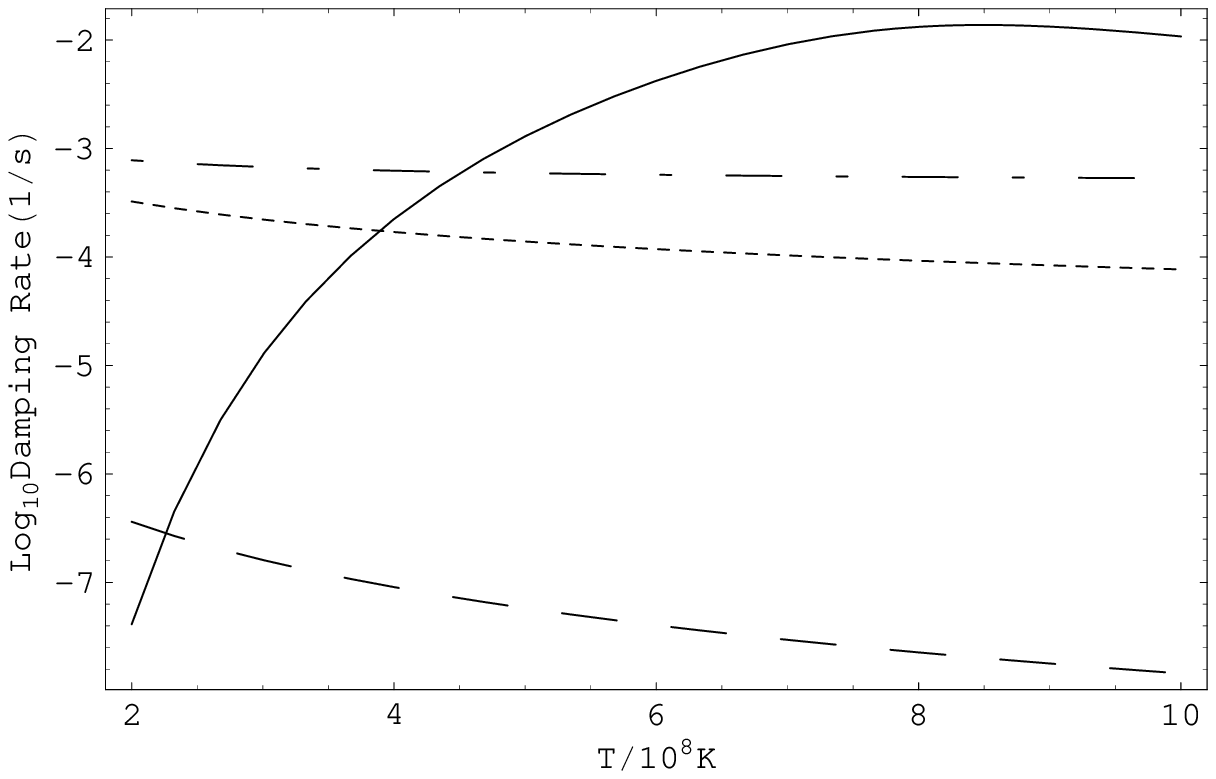}
\figcaption[Fv-l.eps]{The dependence on temperature ($\Te$) of the three contributions to the damping rate $F_v$: core shear viscosity (long dashed), boundary layer viscosity for $B\la 10^9$ Gauss (short dashed) and $B=10^{11}$ Gauss (short and long dashed), and hyperon bulk viscosity (solid). The model chosen has $T_h=2\times 10^9$ K, $\Sc_n=\Sc_s=0.2$, and $\Omt=0.30$.\label{Fv}}
\end{figure}

\begin{figure}
\plotone{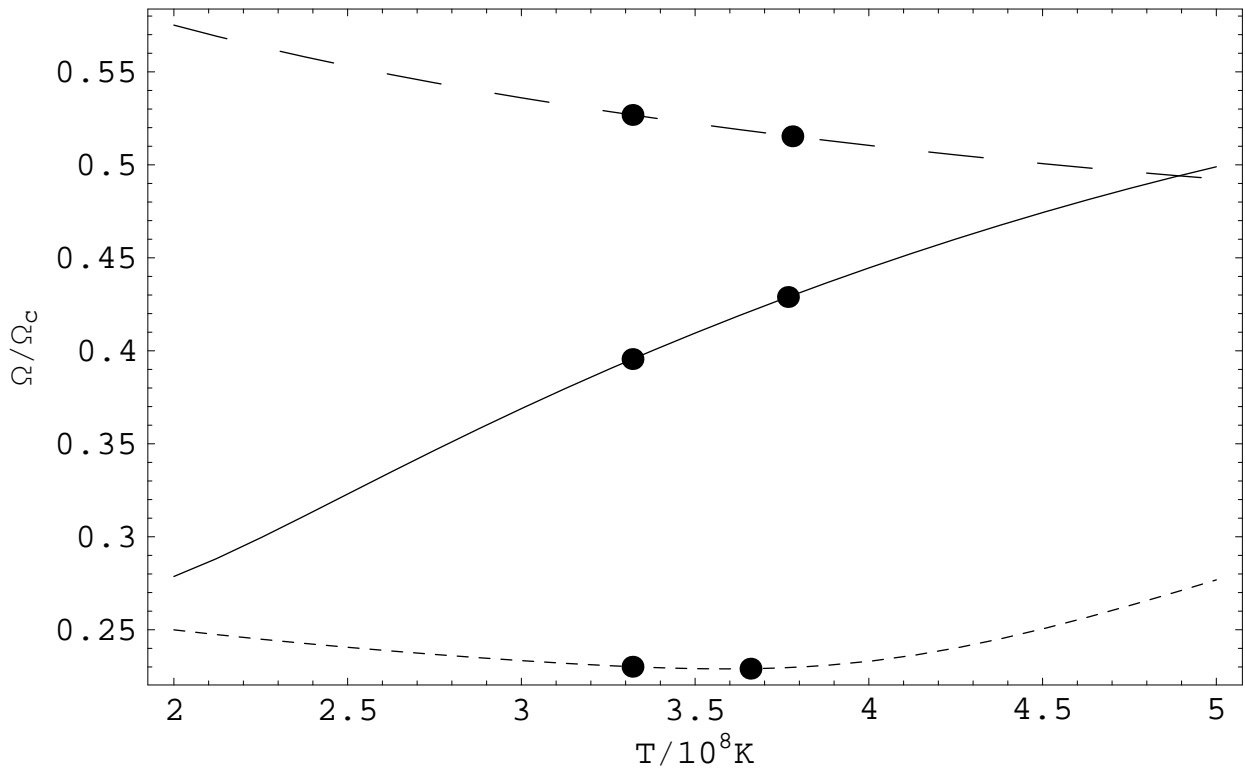}
\figcaption[Omt-l.eps]{The relation bewtween $\Omt=\Om/\Om_c$ and $\Te$ on the critical curve defined by $F_g=F_v$ is shown for the choices (a) $T_h=3\times 10^9$ K, $\Sc_{ns}=1.0$ (long dashed), (b) $T_h=3\times 10^9$ K, $\Sc_{ns}=0.1$ (short dashed), and (c) $T_h=1\times 10^9$ K, $\Sc_{ns}< 0.3$ (solid). The magnetic boundary-layer viscosity is assumed negligible. Also shown are the initial and equilibrium states (at the higher temperature on each curve).\label{Omt}}  
\end{figure}

\end{document}